\documentclass{aa}
\usepackage{upgreek}
\usepackage{graphicx}
\usepackage[varg]{txfonts}
\usepackage{natbib}
\usepackage{float}
\usepackage{soul}
\newlength{\linwx}
\usepackage{color}

\begin{document}

\title{
Influence of grain sizes and composition on the contraction rates of planetary envelopes and on planetary migration}


\author{
Bertram Bitsch \inst{1} and Sofia Savvidou\inst{1}
}
\offprints{B. Bitsch,\\ \email{bitsch@mpia.de}}
\institute{Max-Planck-Institut f\"ur Astronomie, K\"onigstuhl 17, 69117 Heidelberg, Germany
}
\abstract{A crucial phase during planetary growth is the migration when the planetary core has been assembled, but the planet did not open a deep gap yet. During this phase the planet is subject to fast type-I migration, which is mostly directed inwards, and the planet can lose a significant fraction of its semi-major axis. The duration of this phase is set by how long the planetary envelope needs to contract until it reaches a mass similar to the mass of the planetary core, which is when runaway gas accretion can set in and the planet can open a deeper gap in the disc, transitioning into the slower type-II migration. This envelope contraction phase depends crucially on the planetary mass and on the opacity inside the planetary envelope. Here we study how different opacity prescriptions influence the envelope contraction time and how this in turn influences how far the planet migrates through the disc. We find within our simulations that the size distribution of the grains as well as the chemical composition of the grains crucially influences how far the planet migrates before it can reach the runaway gas accretion phase. Grain size distributions with larger grain sizes result in less inward migration of the growing planet, due to faster gas accretion enabled by more efficient cooling. In addition, we find that planets forming in water poor environments can contract their envelope faster and thus migrate less, implying that gas giants forming in water poor environments might be located further away from their central star compared to gas giants forming in water rich environments. Future studies of planet formation that aim to investigate the chemical composition of formed gas giants need to take these effects self consistently into account.
}
\authorrunning{Bitsch and Savvidou}\titlerunning{Gas envelope contraction}\maketitle

\section{Introduction}
\label{sec:Introduction}

The formation of planets and planetary systems in the core-accretion scenario is based on the idea that a planetary core forms first, then contracts a planetary envelope and finally transitions into runaway gas accretion \citep{1996Icar..124...62P}. The envelope contraction rates depend strongly on the mass of the planet and the envelope opacity \citep{2000ApJ...537.1013I, 2014A&A...572A.118M, 2015ApJ...800...82P, 2017A&A...606A.146L, 2017ApJ...836..221M}. During the whole growth process, the planet migrates through the disc, first in type-I migration when it is small and then in the type-II fashion once it becomes big and opens a deep gap in the protoplanetary disc (e.g. \citealt{2012ARA&A..50..211K, 2013arXiv1312.4293B}). The migration rates are strongest for planets above several Earth masses, making the planet most vulnerable to inward migration at this stage before it reaches runaway gas accretion and opens a deep gap \citep{2017Icar..285..145C}. The level of inward migration is then crucially determined by the planetary envelope contraction rates.

The study of gas accretion on planetary cores requires in principle high resolution 3D simulations \citep{2009MNRAS.393...49A, 2013ApJ...778...77D, 2013ApJ...779...59G, 2017A&A...606A.146L, 2017MNRAS.471.4662C, 2019A&A...632A.118S}, which are not only computationally expensive, but are also not yet fully understood. Nevertheless, a clear trend from these simulations seems to have emerged, namely that the opacity in the planetary envelope is crucial for the gas accretion rates. In particular, a lower opacity increases the accretion rates \citep{2017A&A...606A.146L, 2019A&A...632A.118S}. This trend has also been observed in simpler 1D approximations \citep{2000ApJ...537.1013I, 2014A&A...572A.118M, 2015ApJ...800...82P}. In particular \citet{2015ApJ...800...82P} have used a power law grain size distribution with a maximum grain size of 1 cm to calculate the opacities in the planetary envelope.

Planetary migration depends strongly on the radial disc profile (e.g. gas surface density, temperature), but also on the disc's opacity \citep{2011MNRAS.410..293P}. In particular the opacity is responsible for the cooling of the disc around the planet, regulating the entropy related corotation torque and horseshoe drag. In the case of a low opacity, the disc cools very quickly, reaching a nearly locally isothermal state, so that the corotation torque is given by the barotropic, non-linear horseshoe drag plus the linear, entropy-related corotation torque. This generally results in a weaker positive contribution to the torque, leading to a faster inward migration\footnote{The total torque is the combination of the negative Lindblad torque and the barotropic and entropy related corotation torque, which can contribute positively.}.

It is known from observations that giant planets occur more frequently around metal rich stars \citep{2004A&A...415.1153S, 2005ApJ...622.1102F, J2010}. In the core accretion scenario this is explained by the larger availability of solid planetary building blocks, which has been shown in many different simulations (e.g. \citealt{2012A&A...541A..97M, 2018MNRAS.474..886N}). In most of these simulations the opacity in the planetary envelope relevant for the gas contraction rates is assumed for simplification to be either constant \citep{2015A&A...582A.112B} or a scaling of the ISM opacities with some factor \citep{2020arXiv200705561E}. 

In addition, most planet formation simulations assume that all chemical elements scale the same with the iron abundance, [Fe/H]. However, observations of stars inside the Milky Way paint a different picture (e.g. \citealt{2018MNRAS.478.4513B}), which can be explained through different elemental production sites, e.g. low mass stars, supernovae \citep{1957RvMP...29..547B}. \citet{2020A&A...633A..10B} derived the water content of formed super-Earths depending on the underlying stellar abundance and found that stars with super-solar [Fe/H] should host water poor super-Earths, while stars with sub-solar [Fe/H] should host water worlds, on average. This result is based on the assumption that the host star metallicities are a proxy for the chemical composition of the planet forming disc.

In this work we investigate the influence of grain size distributions, calculated self consistently from the disc properties \citep{2011A&A...525A..11B}, and their chemical composition on the opacities in protoplanetary discs and how this influences the envelope contraction rates using a simple recipe\footnote{This simple recipe does not include effects happening {\it inside} the planetary envelope in detail and we discuss the limitations in section~\ref{sec:methods}.} \citep{2000ApJ...537.1013I} and the planetary migration rates following the type-I torque formula of \citet{2011MNRAS.410..293P}.

Our work is structured as follows. In section~\ref{sec:methods} we discuss our disc model, the opacity prescription, the grain size distribution, the envelope contraction rates and the planet migration rates. In section~\ref{sec:growmig} we show how the different grain size distributions and compositions influence planetary growth and migration. We then discuss our results in section~\ref{sec:disc} and conclude in section~\ref{sec:conclude}.

\section{Methods}
\label{sec:methods}

In this section we discuss briefly the methods that we are using in this work. In particular we discuss our disc model, the opacities with their chemical composition, the grain size distribution, the envelope contraction rate and type-I migration.

\subsection{Disc model}

The disc model follows a simplistic power-law in surface density and temperature. With this model, we can change more easily parameters (e.g. viscosity, opacity) compared to very advanced self-consistent hydrodynamical simulations \citep{2020arXiv200514097S}. The disc surface density is inspired by flaring discs with constant radial gas accretion rate $\dot{M}$ and follows
 \begin{equation}
  \label{eq:disc}
  \Sigma_{\rm g} = 1500 \left( \frac{r}{\rm AU} \right)^{-15/14} \frac{\rm g}{\rm cm^2} 
 \end{equation}
where $r$ is the orbital distance. The temperature can be calculated through the disc's aspect ratio, which follows
 \begin{equation}
  \label{eq:discII}
  H/r = 0.033 \left( \frac{r}{\rm AU} \right)^{2/7} \ .
 \end{equation}
This corresponds to an accretion rate of $\approx$10$^{-7} {\rm M}_\odot/{\rm yr}$ for $\alpha=0.01$. In this simple model we make the assumption that the disc's temperature is entirely dominated by stellar irradiation and not by viscous heating. This also implies that the viscosity does not change the disc structure, in contrast to more complex disc models \citep{2020arXiv200514097S}. Varying the $\alpha$ viscosity value thus only has an influence on the grain size distribution and on the planet migration rates (see below) in our model.

\subsection{Opacities and chemical composition}

In order to account for the chemical composition of the disc material, we include only the major rock and ice forming species. The mixing ratios (by number) of the different species as a function of the elemental number ratios is denoted X/H and corresponds to the abundance of element X compared to hydrogen for solar abundances, which we take from \citet{2009ARA&A..47..481A} and are given as follows: He/H = 0.085; C/H = $2.7\times 10^{-4}$; O/H = $4.9\times 10^{-4}$; Mg/H = $4.0\times 10^{-5}$; Si/H = $3.2\times 10^{-5}$; S/H = $1.3\times 10^{-5}$; Fe/H = $3.2\times 10^{-5}$.

These different elements can combine to different molecular and solid species. We list these species, as well as their condensation temperature and their volume mixing ratios $v_{\rm Y}$ in Table~\ref{tab:species}. More details on the chemical model can be found in \citet{2018MNRAS.479.3690B} and \citet{2020A&A...633A..10B}.

We calculate the opacities with the model from the RADMC-3D code\footnote{\url{http://www.ita.uni-heidelberg.de/~dullemond/software/radmc-3d/}}. For the refractory indices we rely on laboratory experiments \citep{1993ApJS...86..713H, 2008JGRD..11314220W, 1993A&A...279..577P, 1997A&A...327..743H, 2003A&A...408..193J}, where the data is available in several databases\footnote{\url{http://vizier.u-strasbg.fr}, \url{https://atmos.uw.edu/ice_optical_constants/}, \url{https://www.astro.uni-jena.de/Laboratory/OCDB/} }. The chemical mixtures used for the opacities are shown in table~\ref{tab:species}.

\begin{figure}
 \centering
 \includegraphics[scale=0.7]{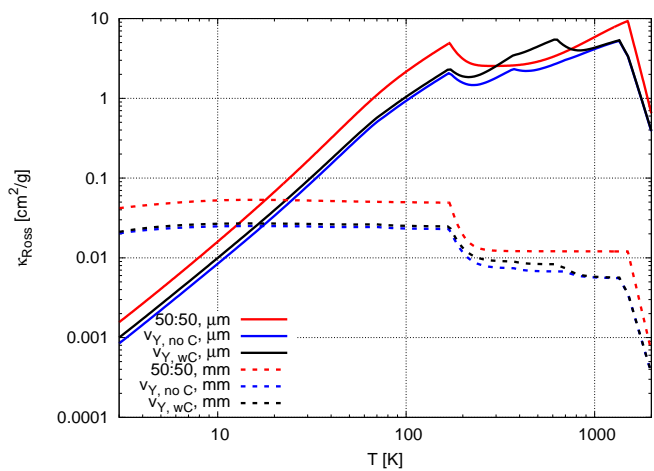}  
 \caption{Rosseland mean opacities $\kappa_{\rm ross}$ as function of temperature for the 3 different compositional assumptions. We use a 50:50 mixture between water and silicates, the complex chemical model without carbon grains and with carbon grains noted as $v_{\rm Y, no C}$ and $v_{\rm Y, w C}$ in table~\ref{tab:species}, respectively. We only show here the opacities for $\mu$m and mm sized dust. A clear difference between the different chemical compositions is visible.
   \label{fig:Rosseland}
   }
\end{figure}

For the calculations of the envelope opacity we rely on the Rosseland mean opacity. The Rosseland mean opacity changes for grains with different sizes and composition, as can be seen in Fig.~\ref{fig:Rosseland}, where we show the Rosseland mean opacity for 3 different compositions and for two single grain sizes. We use here, as in the remaining of our work, a dust-to-gas ratio of 0.01. In our model we use full grain size distributions spanning from sub micron size to cm sizes, and as such also more opacity values as shown in Fig.~\ref{fig:Rosseland}, where we limit ourselves to two sizes to visualize the differences.

In Fig. \ref{fig:Rosseland} it is illustrated how each mean opacity scales with temperature for two different grain sizes, 1 $\mu$m and 1 mm. The wavelength dependent opacities and subsequently the mean opacities depend on the size parameter $x=\frac{2\pi a}{\lambda}$, but also on the refractive index of the given grain species, which is also itself dependent on wavelength \citep{2008Icar..194..368M}. By Wien's law the wavelength is inversely proportional to the temperature. Using the size parameter we find that the regime changes at approximately x = 1 and more specifically at x $\ll$ 1 we have the Rayleigh scattering, whereas at x $\gg$ 1 we have the geometric optics regime. Consequently, if the size of the particle is a lot smaller than the wavelength of the incident radiation, absorption dominates over scattering and the wavelength dependent opacities become independent of grain size. In the case of the larger grain sizes, or when x $\gg$ 1, the opacities become independent of wavelength (and consequently temperature), but depend on the grain size. Most of the regions though lie somewhere in between, which means that calculating the opacity depends on both the grain size with its individual refractive index and the given wavelength or temperature. For more details see \citet{2020arXiv200514097S}.

For the large grains, the opacities are mostly flat, except for transitions at around 170 K, where water ice evaporates, resulting in a drop in the opacity. For the grains with a more complex chemical composition, several ice lines are visible (tab.~\ref{tab:species}). The magnitude of the change of the Rosseland mean opacity at the ice lines depends on the abundances of the evaporating material. As water ice is the most abundant molecule in our model, the change of opacity is largest at the water ice line. The change in opacity visible in the grain composition with carbon (black lines in Fig.~\ref{fig:Rosseland}) at 626K is caused by the pure carbon grains, not present in the other chemical models.

\begin{table*}
\centering
\begin{tabular}{c|c|c|c}
\hline
Species (Y) & $T_{\text{cond}}$ {[}K{]} & $v_{\text{Y, no C}}$ & $v_{\rm Y, w C}$\\ \hline \hline
CO & 20  & 0.45 $\times$ C/H &  0.45 $\times$ C/H  \\[5pt]
CH$_4$ & 30 & 0.45 $\times$ C/H  & 0.25 $\times$ C/H \\[5pt]
CO$_2$ & 70 & 0.1 $\times$ C/H & 0.1 $\times$ C/H \\[5pt]
H$_2$O & 170 & O/H - (3 $\times$ MgSiO$_3$/H + 4 $\times$ Mg$_2$SiO$_4$/H + CO/H \\
& & + 2 $\times$ CO$_2$/H + 3 $\times$ Fe$_2$O$_3$/H + 4 $\times$ Fe$_3$O$_4$/H) & see $v_{\rm Y, no C}$ \\[5pt]
Fe$_3$O$_4$ & 371 & (1/6) $\times$ (Fe/H - S/H) & (1/6) $\times$ (Fe/H - S/H) \\[5pt]
C (carbon grains) & 626 & 0 & 0.2 $\times$ C/H\\[5pt]
FeS & 704 & S/H & S/H \\[5pt]
Mg$_2$SiO$_4$ & 1354 & Mg/H - Si/H & Mg/H - Si/H  \\ [5pt]
Fe$_2$O$_3$ & 1357 & 0.25 $\times$ (Fe/H - S/H) & 0.25 $\times$ (Fe/H - S/H)\\ [5pt]
MgSiO$_3$ & 1500 & Mg/H - 2 $\times$ (Mg/H - Si/H) & Mg/H - 2 $\times$ (Mg/H - Si/H) \\  \hline
\end{tabular}
\caption[Condensation temperatures]{Condensation temperatures and volume mixing ratios of the chemical species. Condensation temperatures for molecules are taken from \citet{2003ApJ...591.1220L}. For Fe$_2$O$_3$ the condensation temperature for pure iron is adopted \citep{2003ApJ...591.1220L}. Volume mixing ratios $v_{\rm Y}$ (i.e. by number) are adopted for the species as a function of disc elemental abundances (see e.g. \citealt{2014ApJ...791L...9M}). We note that the Mg abundance is always larger than the Si abundance. We follow here the different mixing ratios from \citet{2020A&A...633A..10B}.}
\label{tab:species}
\end{table*}

In planet formation theories the chemical abundances of stars is thought to reflect the chemical composition of the natal protoplanetary discs, from which the material stems that planets form from. In most planet formation simulations the host star metallicity, traditionally measured by the iron line, [Fe/H], is a proxy for the total metallicity of the planet forming disc. However, not all elements scale the same with [Fe/H], so that we expect different gradients for different elements. This has been used in \citet{2020A&A...633A..10B} to study the water abundances of super-Earths formed around stars with different composition. We follow here the same approach and list how the abundances of the different elements change with [Fe/H] in table~\ref{tab:abu}.

\begin{table*}
\centering
\begin{tabular}{c|c|c|c|c|c|c|c|c|c}
\hline
[Fe/H] & $\sigma$[Fe/H] & [Si/H] & $\sigma$[Si/H] & [Mg/H] & $\sigma$[Mg/H] & [C/H] & $\sigma$[C/H] & [O/H] & $\sigma$[O/H]\\ \hline \hline
-0.4 & 0.02 & -0.32 & 0.06 & -0.26 & 0.08 & -0.29 & 0.05 & -0.19 & 0.08  \\  
 0.0 & 0.03 &  0.00 & 0.06 &  0.08 & 0.07 &  0.03 & 0.07 &  0.03 & 0.07  \\ 
 0.4 & 0.03 &  0.41 & 0.05 &  0.53 & 0.06 &  0.36 & 0.06 &  0.23 & 0.07  \\  
\hline
\end{tabular}
\caption[Elemental abundances]{Mean stellar abundances as derived from the GALAH catalogue \citep{2018MNRAS.478.4513B} for sun like stars in the solar neighbourhood. Sulfur (not shown) scales in the same way as silicon \citep{2002A&A...390..225C}.}
\label{tab:abu}
\end{table*}

\begin{figure}
 \centering
 \includegraphics[scale=0.7]{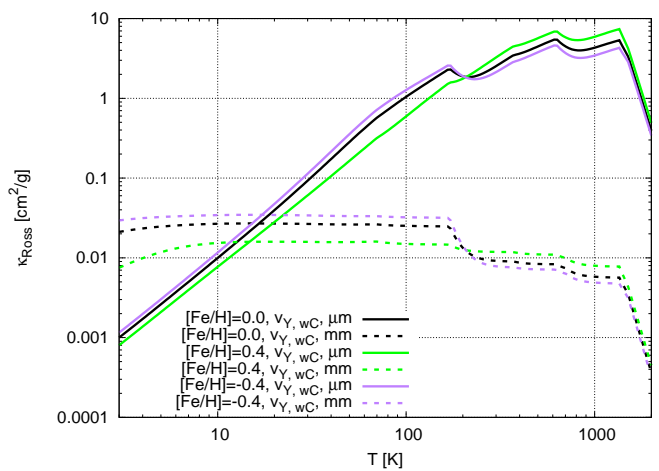}  
 \caption{Rosseland mean opacities $\kappa_{\rm ross}$ as function of temperature for three different mixtures quantified by [Fe/H], see table~\ref{tab:abu}. The opacities in itself are the rescaled to a dust-to-gas ratio of 0.01, as before. The [Fe/H]=0 composition (black line) reflects the same chemical composition as in Fig.~\ref{fig:Rosseland}. By changing the chemical compositions, the changes in opacity at the different ice lines vary.
   \label{fig:RosselandFeH}
   }
\end{figure}

In particular, \citet{2020A&A...633A..10B} found that depending on the C/O ratio of the star, the water abundance inside a protoplanetary disc can vary from about $\approx 50\%$  at [Fe/H]=-0.4 to $\approx 5\%$ at [Fe/H]=0.4. This change in the water abundance, changes the opacity profile around the ice line (Fig.~\ref{fig:RosselandFeH}). In particular for the composition reflecting [Fe/H]=0.4, where the water content is very low, the opacity shows basically no change around the water ice line.

\subsection{Grain size distribution}

Small micrometer sized dust grains can grow in protoplanetary discs through coagulation \citep{2008A&A...480..859B, 2011A&A...525A..11B} and condensation \citep{2013A&A...552A.137R}. The growth of the particles is limited by fragmentation of the grains, which arises when the relative velocities of the individual grains become so large that the grains fragment. This threshold speed, call the fragmentation velocity $u_{\rm f}$ is measured in the laboratory and around $1-10$m/s \citep{2015ApJ...798...34G}. The maximal grain size particles can thus reach is given by
 \begin{equation}
  \label{eq:smax}
  s_{\rm max} \approx \frac{2 \Sigma_{\rm g} u_{\rm f}^2}{\pi \alpha \rho_{\rm s} c_{\rm s}^2} \ ,
 \end{equation}
where $\rho_{\rm s}$ is the density of the pebbles, assumed to be constant and 1.6g/cm$^3$, and $c_{\rm s}$ is the sound speed of the gas.

\begin{figure}
 \centering
 \includegraphics[scale=0.7]{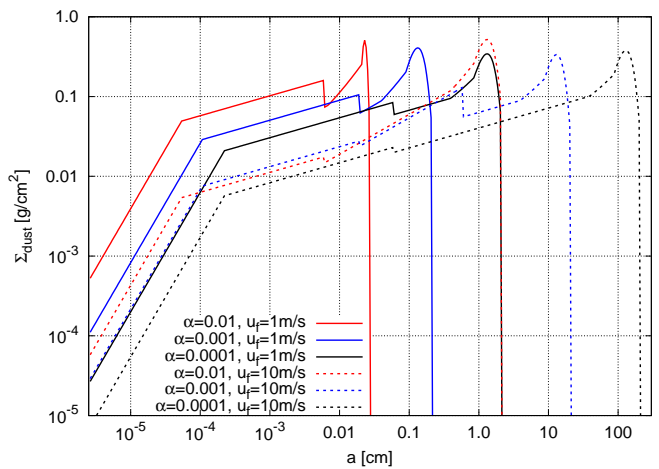} 
 \caption{Dust size distribution for different $\alpha$ values and fragmentation velocities in our disc model at 1 AU. We note that the grain size distribution in itself is independent in our model on the chemical composition of the grains.
   \label{fig:Sigmadust}
   }
\end{figure}

The exact shape of the grain size distribution is a complex interplay between settling, coagulation, cratering and fragmentation. We follow here the outlined grain size distribution recipe from \citet{2011A&A...525A..11B} and show the grain size distribution in our model in Fig.~\ref{fig:Sigmadust}.

The maximal grain size is quadratically proportional to the fragmentation velocity of the grains and inversely proportional to the $\alpha$ parameter (eq.~\ref{eq:smax}). This results in the same maximal grain size for the simulations with $\alpha=10^{-4}; u_{\rm f}=1{\rm m/s}$ and $\alpha=10^{-2}; u_{\rm f}=10{\rm m/s}$. However, the other parts of the grain size distribution are affected, because the settling of the grains is not only a function of the grain size, but also of the turbulence strength, resulting in a different grain size distribution.

In reality, the grains interact with the gas in the protoplanetary disc, which robs them of angular momentum, resulting in an inward drift of pebbles \citep{1977MNRAS.180...57W}. For simplicity we consider here only the Epstein drag. The drift rate of particles depends on their Stokes number, defined as 
\begin{equation}
 \label{eq:Stokes}
 St = \frac{\pi}{2} \frac{\rho_{\rm s} a}{\Sigma_{\rm g}} \ .
\end{equation}
The drift can lead to a pile-up of grains in the inner regions of the disc \citep{2012A&A...539A.148B}. For simplicity we ignore the effects of radial drift, because our work focuses on how the chemical composition of the grains influences the contraction rates of planetary envelopes.

Here the chemical composition would only influence the density of the pebbles, however these changes are of the order unity, so that the radial drift of pebbles is only marginally influenced. As a result, even if pebbles were drifting and piling up in the inner regions of protoplanetary discs, these pile-ups would be nearly identical for all chemical compositions, motivating us to ignore radial drift in this work. In principle a pile-up of dust increases the local dust-to-gas ratio, increasing the opacity and thus prolonging the envelope contraction rates (see below).

In Fig.~\ref{fig:Oparadius} and Fig.~\ref{fig:OpaFEHradius} we show the Rosseland mean opacity as a function of orbital distance for our different assumptions for the grain size distribution and for different chemical compositions. We also show the envelope opacities assumed in \citet{2015A&A...582A.112B}, who used $\kappa_{\rm env} = 0.05 {\rm cm^2/g}$ and of the approach by \citet{2020arXiv200705561E}, who use an ISM opacity (based on micro meter sized grains) scaled by a factor $f=0.003$ motivated by models studying grain growth and settling in planetary atmospheres \citep{2014A&A...572A.118M}. These opacities increase towards the inner disc regions, so towards higher temperatures, in contrast to the other opacity prescriptions. This is caused by the shape of the micro meter opacities, which increase towards higher temperatures, in contrast to the opacities of the grain size distributions which are dominated by the larger grains, where the opacities decrease for higher temperatures (see Fig.~\ref{fig:Rosseland}). The opacities used by \citet{2015A&A...582A.112B} and \citet{2020arXiv200705561E} are much smaller than the ones derived from the full grain size distributions. 

For the opacities derived from grain size distributions, those with larger fragmentation velocities and lower $\alpha$, show the smallest opacity values. This is caused by the fact that grains grow to larger sizes that also contain most of the mass (Fig.~\ref{fig:Sigmadust}, but the larger grains have a lower opacity (Fig.~\ref{fig:Rosseland} and Fig.~\ref{fig:RosselandFeH}) and do not contribute to the opacity significantly. Even though the maximal grain size for $\alpha$=0.01 with $u_{\rm f}=$10.0m/s increases by a factor of 100 for $\alpha=10^{-4}$ with $u_{\rm f}=$10.0m/s, the opacity only slightly decreases. This is caused by the fact that the opacity is dominated by the small grains ($a<100\mu$m) where the surface densities only change by about a factor of 10 (Fig.~\ref{fig:Sigmadust}).

\begin{figure}
 \centering
 \includegraphics[scale=0.7]{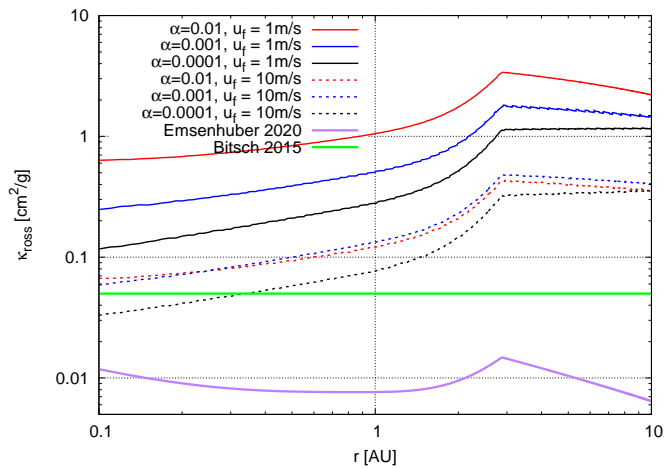} 
 \caption{Rosseland mean opacities $\kappa_{\rm ross}$ as function of orbital distance in our disc model using a 50:50 ratio between water and silicates. The opacities have been calculated using the full grain size distributions. We use different $\alpha$ and fragmentation velocities $u_{\rm f}$ to probe the influence of the grain sizes on the opacities. If the grains grow larger, the opacities are reduced.
   \label{fig:Oparadius}
   }
\end{figure}

Changing the chemical composition of the grains, results in a small change (up to a factor of two) of the opacity (Fig.~\ref{fig:OpaFEHradius}). The differences in the opacities derived for different chemical compositions depend crucially on the disc's temperature and thus on the orbital distance. In particular the opacity is larger in the case of [Fe/H]=0.4 in the inner disc compared to [Fe/H]=-0.4, where this trend flips around the water ice line. This is caused by the different water abundances in these two chemical mixtures, where the [Fe/H]=0.4 model contains less water and thus show a lower opacity.

\begin{figure}
 \centering
 \includegraphics[scale=0.7]{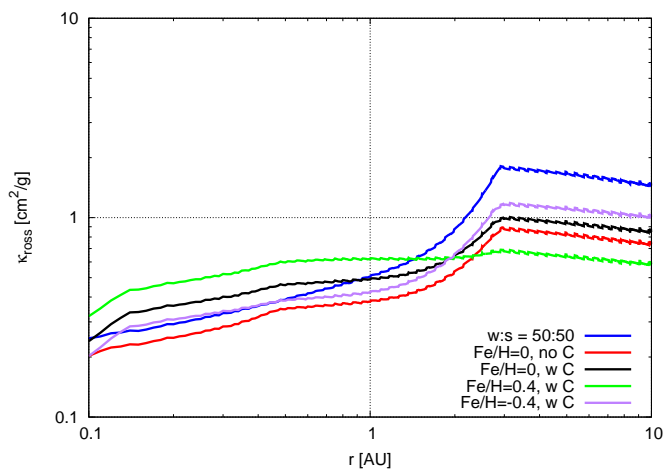} 
 \caption{Rosseland mean opacities $\kappa_{\rm ross}$ as function of orbital distance in our disc model using $\alpha=0.001$ and $u_{\rm f} =1$m/s for all chemical compositions. We vary here the chemical composition of the grains in the disc, following the approach of \citet{2020A&A...633A..10B}. Here [Fe/H] implies only a change of the composition of the grains (table~\ref{tab:abu}), not a change of the dust to gas ratio. We note that in \citet{2020A&A...633A..10B} [Fe/H]=0.4 implies a very low water fraction, while [Fe/H]=-0.4 provides a large water fraction.
   \label{fig:OpaFEHradius}
   }
\end{figure}

\subsection{Envelope contraction rates}

The contraction rates of planetary envelopes has been studied in the past in many different frameworks \citep{2000ApJ...537.1013I, 2009MNRAS.393...49A, 2014ApJ...789L..18O, 2014A&A...572A.118M, 2014ApJ...797...95L, 2015ApJ...811...41L, 2017A&A...606A.146L, 2017MNRAS.471.4662C, 2019A&A...632A.118S}. All of these studies find that the envelope contraction rate depends mainly on the planetary mass and the opacity within the planetary envelope. 

The opacities within the planetary envelope depend on the grain evolution within the planetary envelope \citep{2014ApJ...789L..18O, 2014A&A...572A.118M}. More precisely, the envelope contraction rates depend on the opacity at the radiative-convective boundary inside the planetary atmosphere \citep{2014ApJ...797...95L, 2015ApJ...811...41L}, where dust grains are already evaporated and the opacity is dominated by H- ions. Furthermore, recycling flows can penetrate within the planetary Hill sphere, removing gas before it can cool and contract, potentially preventing gas accretion at close orbits \citep{2017A&A...606A.146L, 2017MNRAS.471.4662C}. Including these effects is critical to understand the contraction rates of planetary envelopes in detail. However, here we want to investigate in a simple way how and if the composition and the inclusion of grain size distributions for the opacity (compared to simple opacity laws) affect the growth and migration of planets. We thus chose a very simple model for the contraction rate, following \citet{2000ApJ...537.1013I}, where the gas contraction rate is given as
\begin{equation}
 \label{eq:Ikoma2000}
 \dot{M}_{\rm gas} = \frac{M_{\rm planet}}{\tau_{\rm KH}} \quad , \quad \tau_{\rm KH} = \left(\frac{M_{\rm core}}{\rm 1 M_{\rm E}} \right)^{-2.5} \left( \frac{\kappa_{\rm ross}}{1 \rm cm^2/g} \right) 10^8 {\rm yr} \ .
\end{equation}
Here $M_{\rm planet}$ is the total planetary mass, $M_{\rm core}$ is the mass of the planetary core, while $\tau_{\rm KH}$ represents the Kelvin-Helmholtz contraction time.

\begin{figure}
 \centering
 \includegraphics[scale=0.7]{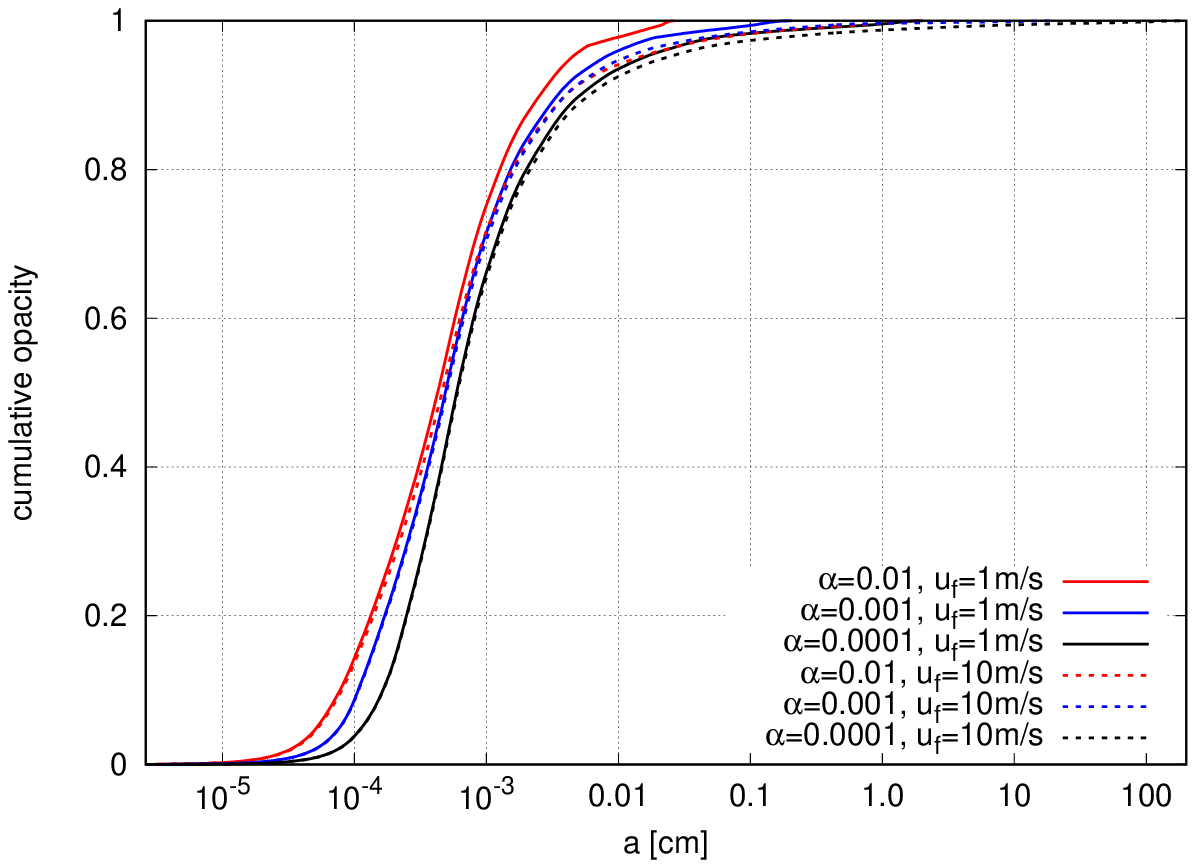} 
 \includegraphics[scale=0.7]{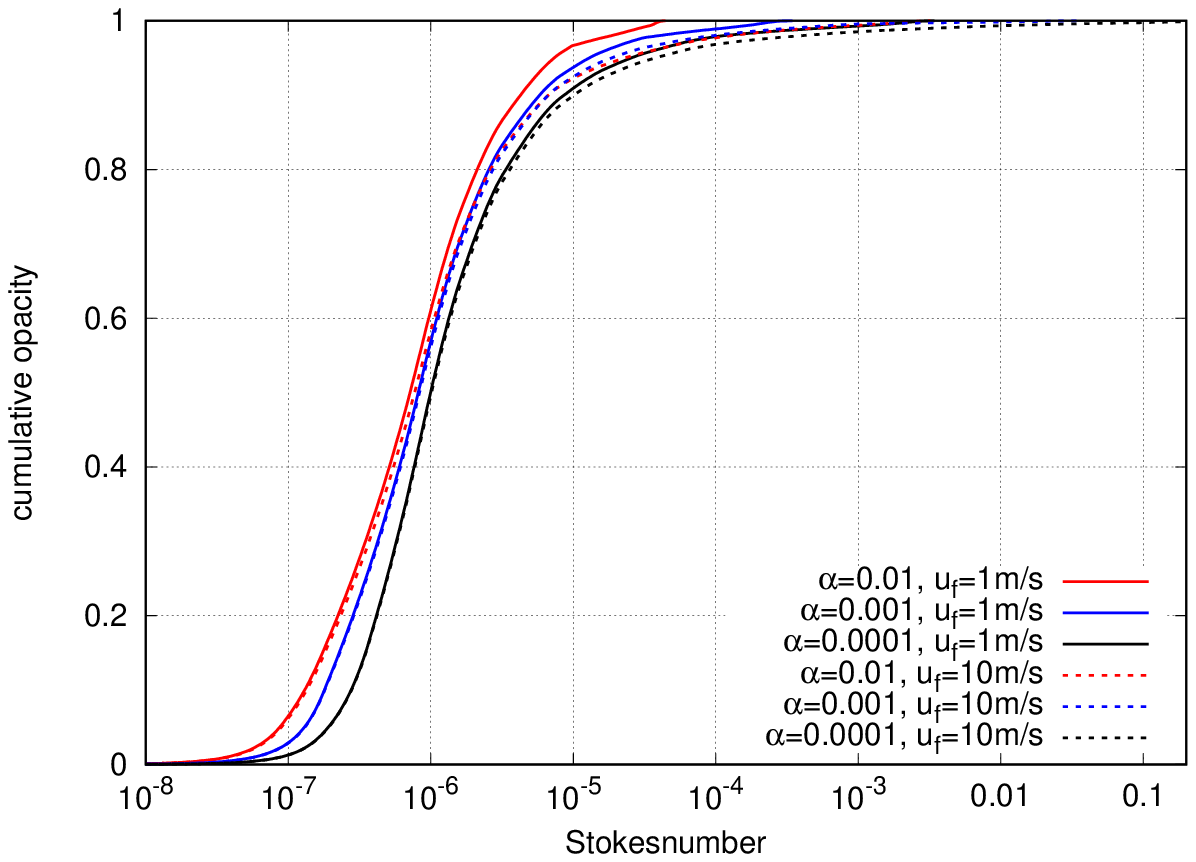}  
 \caption{Cumulative Rosseland mean opacities in discs at 1 AU with different $\alpha$ viscosity parameters and for different fragmentation velocities. The main contribution to the opacities is done by the small particles below 10-100 micron or when converted into Stokes numbers from particles with Stokes numbers less than $10^{-4}$, where particles are perfectly coupled to the gas and are thus not blocked in the pressure bump generated by the planet.
   \label{fig:Opacum}
   }
\end{figure}

As the opacity changes with grain size (Fig.~\ref{fig:Rosseland}) and as the grain size distributions vary with $\alpha$ and $u_{\rm f}$, we show in Fig.~\ref{fig:Opacum} the cumulative Rosseland mean opacities at 1 AU in our disc model for different $\alpha$ and fragmentation velocities as function of grain size (top) and Stokes number (bottom). It is important to note that the absolute value of opacity decreases when the grain sizes become larger as more and more mass is transfered to large grains with low opacities.

Interestingly the shape of the cumulative opacities is very similar in all cases. The reason for that is related to the scaling of the opacities with grain size. Once the grains have reached mm in size, the opacities decrease linearly for larger grains. This is related to the size parameter $x$, which determines the opacity value for large grains (see above) and results in a linear scaling with grain size. As a result, the opacities are dominated by the small grains, even though they contain less mass, compared to the large grains. For all our levels of turbulence strength and fragmentation velocity, 90$\%$ of the opacities are contributed from grains smaller than 100 micron.

As a planet grows, it starts to carve a small gap in its natal protoplanetary disc and generates a dust trap exterior to its orbit \citep{2004A&A...425L...9P, 2006A&A...453.1129P}. This dust trap is caused by a small pressure bump that the planet generates, which can halt the inward flowing pebbles \citep{2012A&A...546A..18M, 2014A&A...572A..35L, 2018arXiv180102341B, 2018A&A...615A.110A}. In the pebble accretion scenario, this is refered to as the pebble isolation mass, at which the planet can slowly start to contract its envelope. 

The pebble isolation mass in itself depends, among other things, on the turbulence strength and on the size of the pebbles in the disc. Planets are normally very inefficient to block pebbles with Stokes numbers smaller than the level of turbulence \citep{2016A&A...585A..35P, 2018arXiv180102341B}. This implies that planets would be very inefficient in blocking pebbles with Stokes numbers lower than $10^{-4}$. However, grains with Stokes numbers larger than $10^{-4}$ barely contribute to the opacity (Fig.~\ref{fig:Opacum}). We thus ignore the reduction of the opacity due the blockage of larger grains exterior to the planet in our approach.

On the other hand, the large grains inside the pressure bump can collide and fragment, producing large numbers of small grains, which can diffuse through into the inner disc, where they can be accreted by the planet \citep{2020ApJ...896..135C}. This effect can increase the opacity inside the planetary envelope, delaying effective gas accretion. However, the increased opacity in itself would still depend on the underlying composition of the grains (Fig.~\ref{fig:Rosseland} and Fig.~\ref{fig:RosselandFeH}).

\begin{figure}
 \centering
 \includegraphics[scale=0.7]{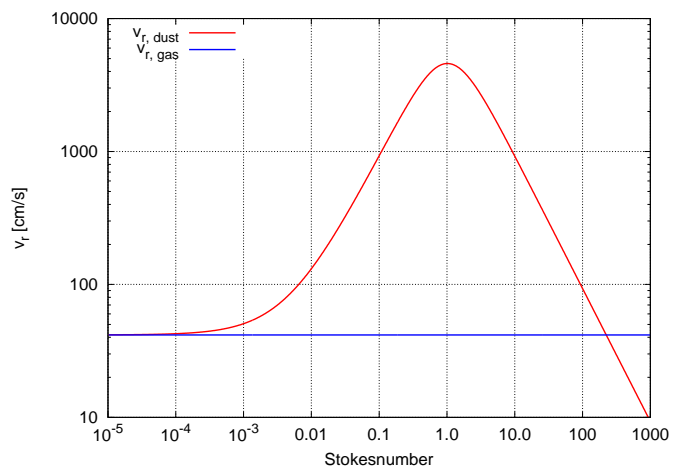}  
 \caption{Radial drift speed of particles in comparison to the gas velocities at 1 AU. Small particles are coupled to the gas and drift inwards with the gas. These particles (St$<10^{-4}$) carry most of the opacity (Fig.~\ref{fig:Opacum}), contributing to the envelope contraction rates.
   \label{fig:drift}
   }
\end{figure}

In addition it is important to note that when the planet accretes gas from its surroundings it would accrete the particles that are coupled to the gas. We can calculate particles of which size are coupled to the gas by evaluating the drift speed of the particles following \citet{2008A&A...480..859B}:
\begin{equation}
 v_{\rm d,rad,tot} = v_{\rm r, d} + \frac{v_{\rm r,gas}}{1+St^2} \ .
\end{equation}
The radial speed of the gas $v_{\rm r,gas}$ in an $\alpha$ disc is estimated by \citet{Takeuchi2002} as
\begin{equation}
 v_{\rm r,gas} = - 3 \alpha \frac{c_{\rm s}^2}{v_{\rm K}} \left( \frac{3}{2} - \alpha_\Sigma \right) \ .
\end{equation}
The quantity $v_{\rm r,d}$ that describes the radial drift of individual dust particles is given by \citet{1977MNRAS.180...57W} as
\begin{equation}
 v_{\rm r, d} = - \frac{2 \Delta v}{St + 1/St} \ ,
\end{equation}
where $\Delta v$ is given as
\begin{equation}
 \Delta v = \eta v_{\rm K}
\end{equation}
with
\begin{equation}
 \label{eq:eta}
 \eta = - \frac{1}{2} \left(\frac{H}{r}\right)^2 \frac{\partial \ln P}{\partial \ln r} \ .
\end{equation}
Here $\frac{\partial \ln P}{\partial \ln r}$ represents the radial pressure gradient. We plot the velocities of the dust grains as function of their Stokes number in Fig.~\ref{fig:drift}. Particles with Stokes numbers less than $10^{-4}$ are coupled to the gas and follow the motion of the gas. At the same time, the particles coupled to the gas carry most of the opacity, while most of the mass is carried in particles that are rapidly drifting inwards.

Our model is based on the assumption that the opacity in the midplane of the protoplanetary disc reflects the opacity in the planetary envelope. This assumption is justified, because the opacity is dominated by the small grains, which are perfectly coupled to the gas and are thus accreted by the planet. In addition we assume that the grains do not evaporate inside the planetary atmosphere.

\begin{figure}
 \centering
 \includegraphics[scale=0.7]{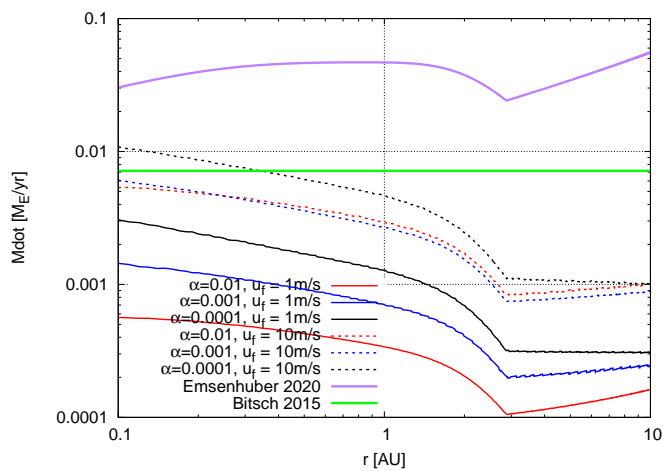} 
 \caption{Gas accretion rates on planetary cores with initially 20 Earth masses implanted at different positions in the protoplanetary disc. The opacity needed for envelope contraction (eq.~\ref{eq:Ikoma2000}) follows directly the midplane opacity of the disc (Fig.~\ref{fig:Oparadius}). A lower opacity increases the accretion rate.
   \label{fig:Ikoma20}
   }
\end{figure}

\begin{figure}
 \centering
 \includegraphics[scale=0.7]{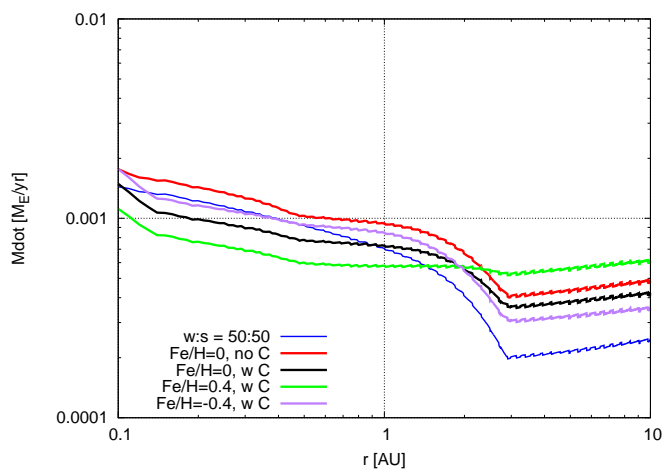} 
 \caption{Gas accretion rates on planetary cores with initially 20 Earth masses implanted at different positions in the protoplanetary disc. The opacity needed for envelope contraction (eq.~\ref{eq:Ikoma2000}) follows directly the midplane opacity of the disc (Fig.~\ref{fig:OpaFEHradius}), where the grain size distribution was derived using $\alpha$=0.001 and $u_{\rm f}=1$m/s. The inclusion of carbon grains and also the water-to-silicate ratio matters for the opacities and thus envelope contraction rates.
   \label{fig:Ikoma20FEH}
   }
\end{figure}

Using this assumption, we calculate the envelope contraction rates $\dot{M}$ for planets with a core of 20 Earth masses for different opacity environments in Fig.~\ref{fig:Ikoma20} and Fig.~\ref{fig:Ikoma20FEH}. In addition to the full grain size distributions we show the contraction rates for the constant envelope opacity approach of \citet{2015A&A...582A.112B}, who used $\kappa_{\rm env} = 0.05 {\rm cm^2/g}$ and of the approach by \citet{2020arXiv200705561E}, who use and ISM opacities scaled by a factor $f=0.003$. For the ISM opacity we use directly the opacities derived from pure micro meter sized grains.

The gas accretion rates shown in Fig.~\ref{fig:Ikoma20} reflect the expectations from the radial opacity distribution (Fig.~\ref{fig:Oparadius}), namely that lower opacities allow larger accretion rates. In particular, the accretion rates derived from the envelope opacities used in \citet{2015A&A...582A.112B} and \citet{2020arXiv200705561E} give the highest accretion rates. In these cases the planet would reach 40 Earth masses, where runaway gas accretion would start, in less than 1000 years. However, this calculation does not take into account how much gas is actually available locally in the disc at the planet's position and how much gas can be provided by the disc to the planet due to viscosity.

In contrast, the accretion rates derived from the opacities of the full grain size distributions are much lower, especially for small fragmentation velocities. In this case, around 10kyr are needed for the planet to grow to the runaway gas regime.

In Fig.~\ref{fig:Ikoma20FEH} we show the envelope contraction rates for planets with different envelope opacities, derived from grain size distributions using $\alpha$=0.001 and $u_{\rm f}=1$m/s and different chemical compositions. The differences between the different chemical compositions are of the order of a few, but are all within the same order of magnitude. The gas accretion rates are reflected by the opacities (Fig.~\ref{fig:OpaFEHradius}), which flip for the different compositions at the water ice line, resulting in higher accretion rates for discs with a lower water ice content, because a larger water fraction results in a larger opacity at $T<$170K (Fig.~\ref{fig:RosselandFeH}).

\subsection{Planet migration}

Planets interact gravitationally with their natal protoplanetary disc, which results in an exchange of angular momentum between planet and disc, leading to planetary migration \citep{1986Icar...67..164W}. Small mass planets migrate in type-I migration, while large gap opening planets migrate in the type-II fashion. Our initial planetary mass is 20 Earth masses, which should only perturb the disc slightly \citep{2009A&A...506..971K, 2014MNRAS.440..683L} for high viscosities, but might already open deeper gaps at low viscosities \citep{2008ApJ...672.1054B, 2013A&A...550A..52B}. We thus include in our model a type-I migration approximation including a transition to type-II migration, following the approach of \citet{2018arXiv180511101K}. We briefly outline our approach for planet migration.

In the type-I regime, we follow the torque formalism of \citet{2011MNRAS.410..293P}, which includes prescriptions for the Lindblad torque, the barotropic and entropy related corotation torque. The recent torque formalism by \citet{2017MNRAS.471.4917J} expands on the approach by \citet{2011MNRAS.410..293P} by adapting the torque formula to the new results of 3D hydrodynamical simulations in contrast to the 2D simulations by \citet{2011MNRAS.410..293P}. However, the differences in the final planetary positions seem quite small \citep{2020arXiv200400874B}. 

The torque formalism of \citet{2011MNRAS.410..293P} requires to specify an opacity responsible for the cooling around the planet, needed to calculate the entropy driven corotation torque. In our simulations we use the same opacities for the calculation of the planetary envelope and for the migration rates. Exceptions are only the case of the \citet{2015A&A...582A.112B} and \citet{2020arXiv200705561E} envelope opacities. In this case we use, as in both their models, the opacity provided by pure $\mu$m sized grains.

Planets that start to accrete gas efficiently, start to open deep gaps in the protoplanetary disc, indicating a change of the migration regime. \citet{2018arXiv180511101K} relate the type-II migration time-scale to the type-I migration time-scale (which we calculate as explained above) in the following way
\begin{equation}
\label{eq:migII}
 \tau_{\rm mig II} = \frac{\Sigma_{\rm up}}{\Sigma_{\rm min}} \tau_{\rm mig I} \ ,
\end{equation}
where $\Sigma_{\rm up}$ corresponds to the unperturbed gas surface density and $\Sigma_{\rm min}$ to the minimal gas surface density at the bottom of the gap generated by the planet. The ratio $\Sigma_{\rm up}/\Sigma_{\rm min}$ can be expressed through \citep{2013ApJ...769...41D, 2014ApJ...782...88F, 2015MNRAS.448..994K}
\begin{equation}
\label{eq:Kgapopen}
 \frac{\Sigma_{\rm up}}{\Sigma_{\rm min}} = 1 + 0.04 K_{\rm mig} \ ,
\end{equation}
where
\begin{equation}
 K_{\rm mig} = \left( \frac{M_{\rm P}}{{\rm M}_\odot} \right)^2 \left( \frac{H}{r} \right)^{-5} \alpha_{\rm mig}^{-1} \ .
\end{equation}
The transition to the pure type-II migration regime can be changed due to gas accretion by growing planets \citep{Bergez20}. We use here for the migration rate the same viscosities as for the grain size distributions. We note that at high viscosities, the entropy driven corotation torque could operate and drive outward migration in certain regions of the disc\citep{2015A&A...575A..28B}. Applying the type-II migration rate from \citet{2018arXiv180511101K} in the case of high viscosity could lead to an nonphysical outward migration in the type-II regime. However, the profiles of our disc are such that type-I migration, and consequently also type-II migration, is always directed inwards.

\begin{figure}
 \centering
 \includegraphics[scale=0.7]{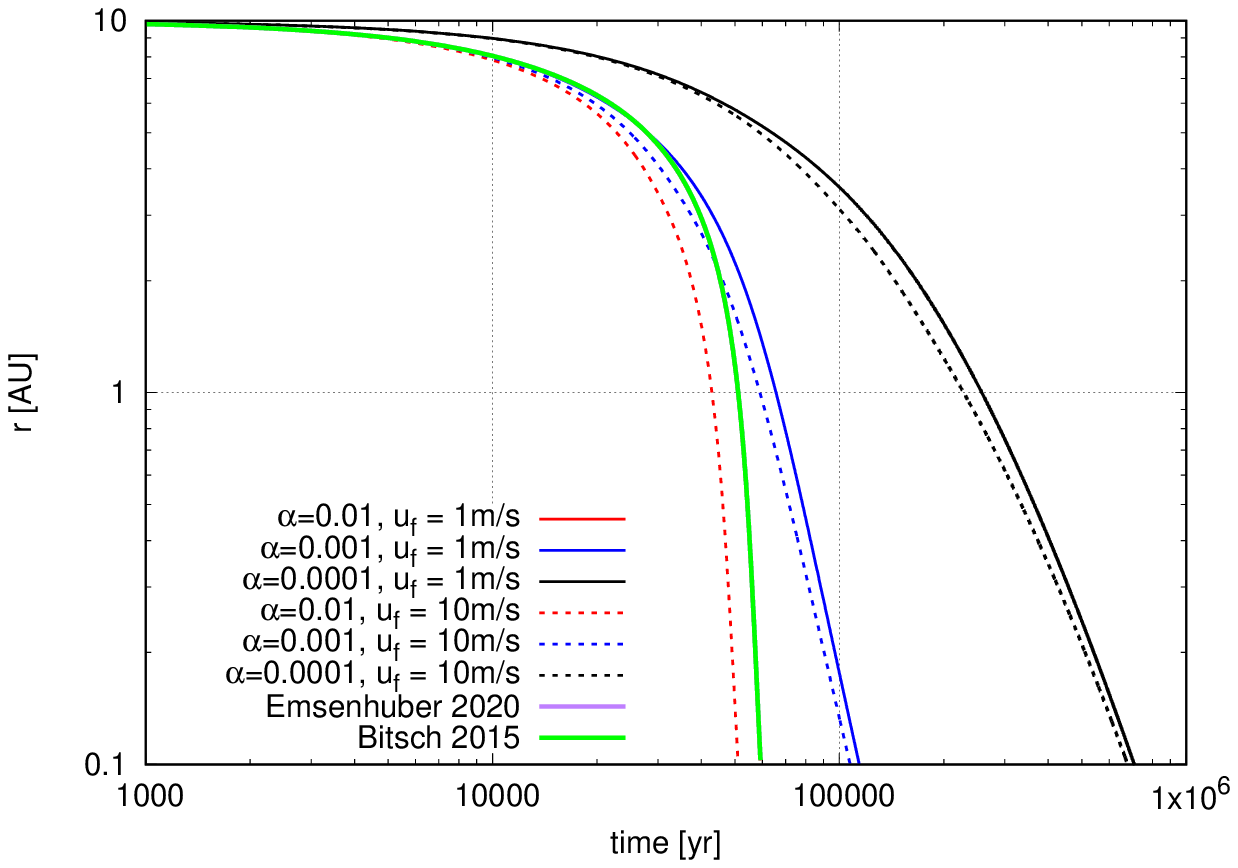} 
 \includegraphics[scale=0.7]{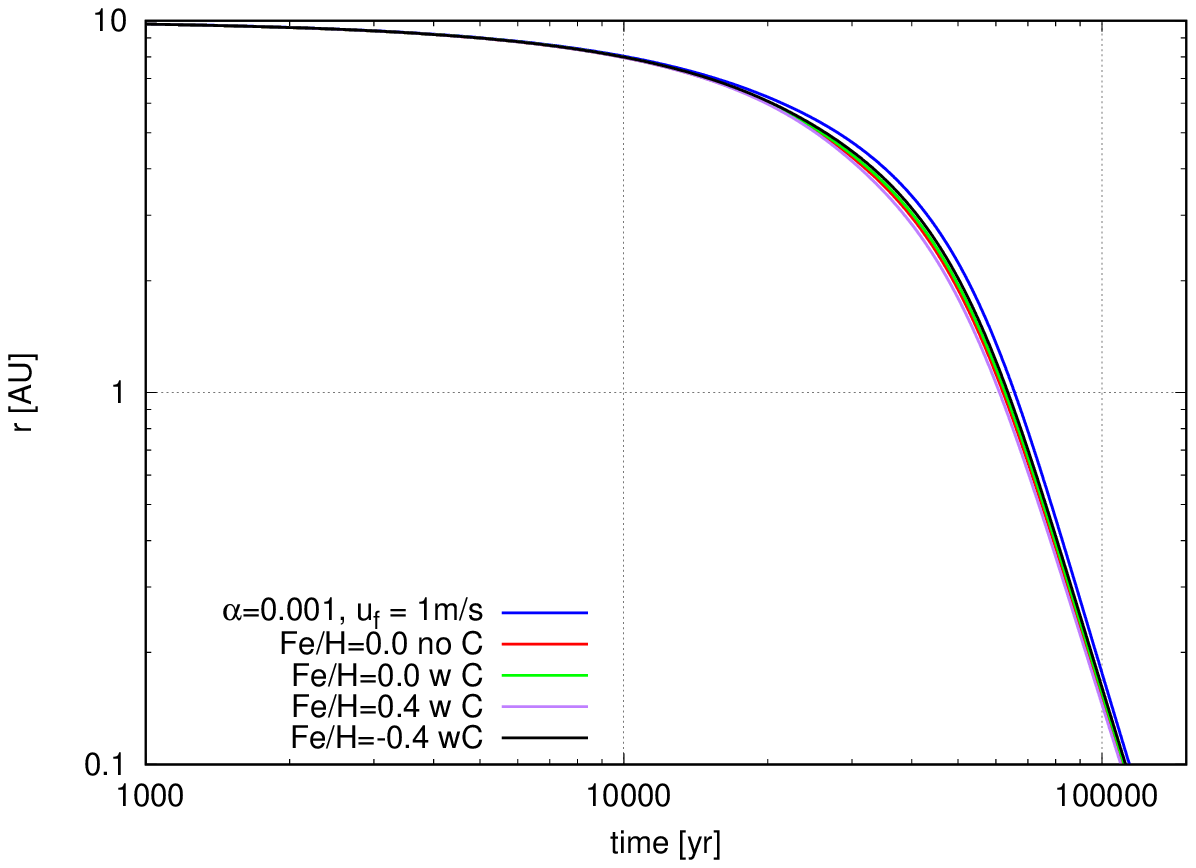}  
 \caption{Change of semi-major axis as function of time for a planet with a constant mass of 20 Earth masses embedded in our disc model at initially 10 AU for different prescriptions of $\alpha$ (top) and for different chemical compositions for the opacity (bottom). All simulations in the bottom plot feature $\alpha=0.001$ and $u_{\rm f}=$1m/s. The two opacity descriptions using \citet{2015A&A...582A.112B} and \citet{2020arXiv200705561E} for the envelope opacity, use pure micro meter grain opacities for migration, resulting in the identical migration patterns.
   \label{fig:rtime}
   }
\end{figure}

In Fig.~\ref{fig:rtime} we display the evolution of the planetary semi-major axis as function of time for a planet with a fixed mass of 20 Earth masses starting at 10 AU from its host star. We show curves following different prescriptions of the disc's opacity and stop the simulations once the plant reaches 0.1 AU. The viscosity used to calculate the opacities is the same opacity used for planet migration. This leads to the expected behavior that planets migrating in low viscosity discs migrate less, as they perturb the disc more, leading to a reduced migration rate (eq.~\ref{eq:migII}). The curves for the opacity model following \citet{2015A&A...582A.112B} and \citet{2020arXiv200705561E} are on top of each other, as they both use the same opacity for migration and only differ for the envelope opacities.

In contrast, the differences between the different opacity prescriptions are very small. This is related to the fact that the opacity only plays a role to decide if the torque is computed via the linear corotation torque or the horseshoe drag. The other parts of the torque (the Lindblad torque and the barotropic corotation torque) are unaffected by the opacity. In addition the opacities for our different chemical compositions are quite similar (Fig.~\ref{fig:OpaFEHradius}), so that their influence on the migration rates is quite small. In the next section, we combine the accretion and migration in one simple model to investigate the influence of the different composition.

\section{Growth versus migration}
\label{sec:growmig}

We now combine our different ingredients (opacities, envelope contraction and planet migration) in one model to investigate how the different opacities influence the final position of the growing planets. As before, we assume that a planetary core with a mass of 20 Earth masses has already been formed. We then calculate how much the planet migrates until it reaches the runaway gas accretion phase, defined as $M_{\rm core} = M_{\rm env}$, when we stop our simulations. We also stop our simulations if the planet reaches the disc's inner edge at 0.1 AU.

\begin{figure}
 \centering
 \includegraphics[scale=0.7]{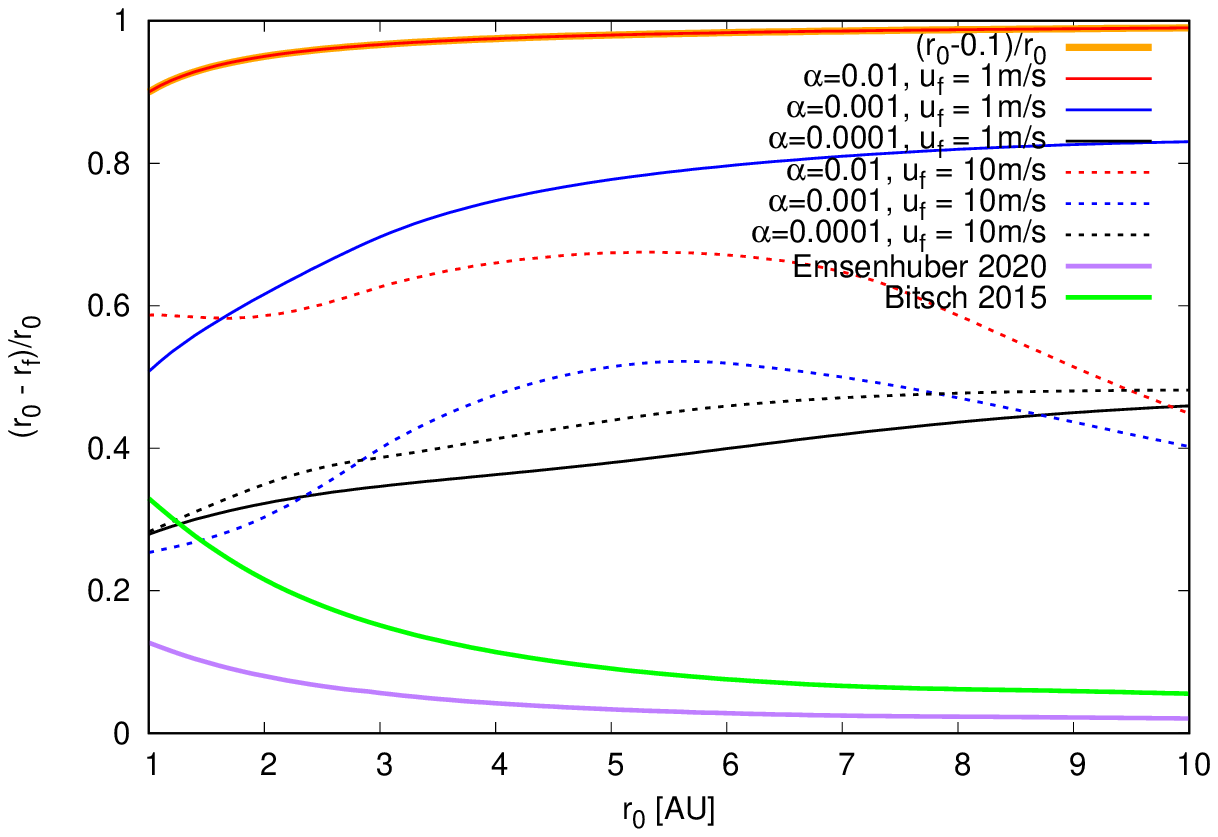} 
 \includegraphics[scale=0.7]{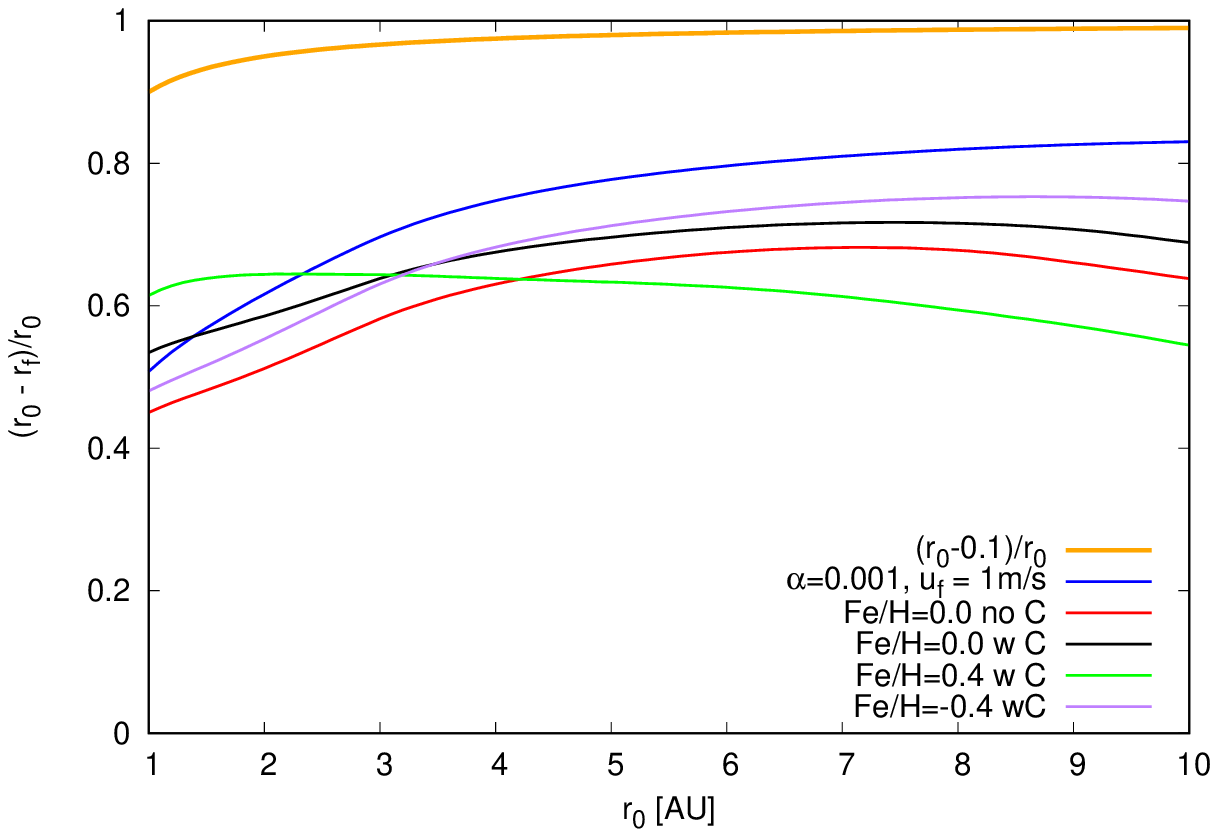}  
 \caption{Change of semi-major axis relative to the initial planetary positions for planetary cores with initially 20 Earth masses growing by gas accretion until they reach $M_{\rm core}=M_{\rm env}$, defined as the onset of runaway gas accretion. We show in the top panel the change of semi-major axis for the different grain size distributions and in the bottom panel we show the change of semi-major axis for different chemical assumptions, where all simulations in the bottom plot feature $\alpha=0.001$ and $u_{\rm f}=1$m/s. A value of $(r_0-r_{\rm f})/r_0=0.0$ implies that the planet basically does not migrate while it is accreting gas. The orange line marks the value if the planet reaches the inner edge of the disc at 0.1 AU.
   \label{fig:r0rf}
   }
\end{figure}

In Fig.~\ref{fig:r0rf} we show the change of the final semi-major axis compared to the initial semi-major axis of planets accreting gas and migrating in different opacity environments. The orange line depicts the maximal change of the semi-major axis for a planet, e.g. a planet starting at 1.0 AU, can only lose 90\% of its initial semi-major axis, because the inner disc edge is situated at 0.1 AU. As the inner edge is always at 0.1 AU, the maximum loss of semi-major axis of the planet expressed in percent increases if the planet has initially a larger semi-major axis (e.g. a planet starting at 10 AU migrating down to 0.1 AU lost 99\% of its initial semi-major axis).

In the top panel of Fig.~\ref{fig:r0rf} we show the loss of semi-major axis for our simulations where the opacity is calculated for full grain size distributions with different levels of viscosity and with different fragmentation velocities. As before, we use the same opacities for the contraction rates and for the migration. For lower viscosity and higher fragmentation velocities, we expect larger grains (Fig.~\ref{fig:Sigmadust}) and thus lower opacities (Fig.~\ref{fig:Oparadius}). As a consequence the planet contracts the envelope faster and the planet thus loses less of its initial semi-major axis.

For $\alpha=0.01$ and $u_{\rm f}$=1m/s, the planet always migrates all the way to the inner edge of the protoplanetary disc. However, if $u_{\rm f}=$10m/s, the planet migrates much less, because of the reduced opacities which allows a faster envelope contraction.

For fragmentation velocities of $u_{\rm f}=10.0$m/s, we see that the trend with less loss of semi-major axis with decreasing viscosity is not entirely linear. This is caused by the fact that the larger grain sizes lead to lower opacities, which in turn increase the thermal diffusivity of the disc. A larger thermal diffusivity results in a basically isothermal behavior of the disc regarding planet migration \citep{2011MNRAS.410..293P}. In this case, the corotation torque is dominated by the linear entropy torque, which results in faster inward migration compared to the entropy related horseshoe drag. We thus observe that even though the envelope contraction is faster for $\alpha=10^{-4}$, planets lose slightly more of their initial semi-major axis compared to the case of $\alpha=0.01$ for $u_{\rm f}$=10m/s.

For the simulations using the opacities used in the planet formation models of \citet{2015A&A...582A.112B} and \citet{2020arXiv200705561E}, we observe only a very small inward migration, due to the fast envelope contraction. At the same time the high opacities of the disc\footnote{We remind that we use for these cases the pure $\mu$m opacities to calculate the migration rate.}, allow the entropy related horseshoe drag to operate, which slows down the inward migration. In the end, the difference in semi-major axis loss between these two sets of simulations is quite similar.

Our results clearly indicate that a change in the opacity of the planetary envelope has large consequences for the final position of a gas envelope contracting planet. In particular models using a low opacity in the planetary envelope show basically no loss of semi-major axis during the contraction phase compared to models where the opacity is derived via full grain size distributions.

In the bottom panel of Fig.~\ref{fig:r0rf} we show the loss of semi-major axis for gas contracting planets embedded in discs where the opacities are calculated by full grain size distributions with $\alpha$=0.001 and $u_{\rm f}=1$m/s, but for different grain compositions\footnote{For $\alpha=0.01$, the chemical composition makes no difference, because the planets always migrate to the inner disc edge before they reach the runaway phase in our model.}. The envelope contraction rates differed by a factor of a few for the different grain compositions (Fig.~\ref{fig:Ikoma20FEH}), which also leads to a difference in the loss of semi-major axis for the different cases. 

For planets forming within the inner few AU, we see only small differences in the final semi-major axes of the planet, which is related to the fact that nearly all planets migrate to the inner edge of the protoplanetary disc. The difference between the planets starting in discs with different chemical compositions becomes apparent in the outer regions of the protoplanetary disc.

Here again the differences in opacities have two effects: (i) a change in the envelope contraction rate and (ii) a change in the planet's migration rate. In particular the planets growing and migrating in a disc, where the chemical composition is made of a 50:50 mixture between silicate and water, lose most of their initial semi-major axis. This is caused by the fact that the opacities in these discs are the highest due to the largest water ice fraction.

The differences in final position for planets forming in discs where the opacities are calculated through a more complex chemical composition show small difference if carbon grains are included or not. If pure carbon grains are included, the overall opacities are higher due the large opacities of the carbon grains, leading to slightly longer contraction times and thus more inward migration.

The simulations utilizing an opacity derived from grains with a composition reflecting [Fe/H]=0.4 have the lowest opacity in the outer regions (Fig.~\ref{fig:OpaFEHradius}), allowing thus the fastest contraction rates, resulting in the least loss of semi-major axis. The difference in the final semi-major axis compared to planets forming in discs with a composition reflecting [Fe/H]=-0.4 can be up to a factor of 2.

\subsection{Dependency on the initial core mass}

\begin{figure}
 \centering
 \includegraphics[scale=0.7]{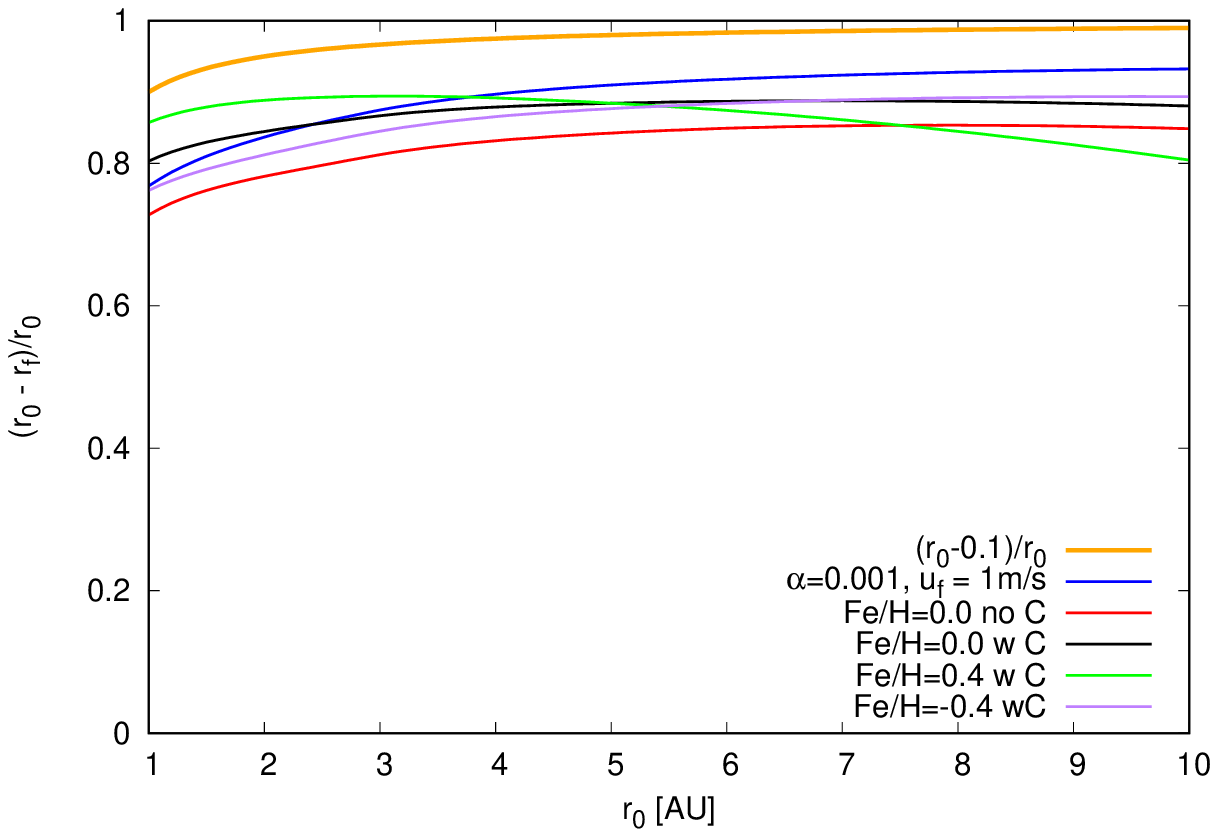} 
 \includegraphics[scale=0.7]{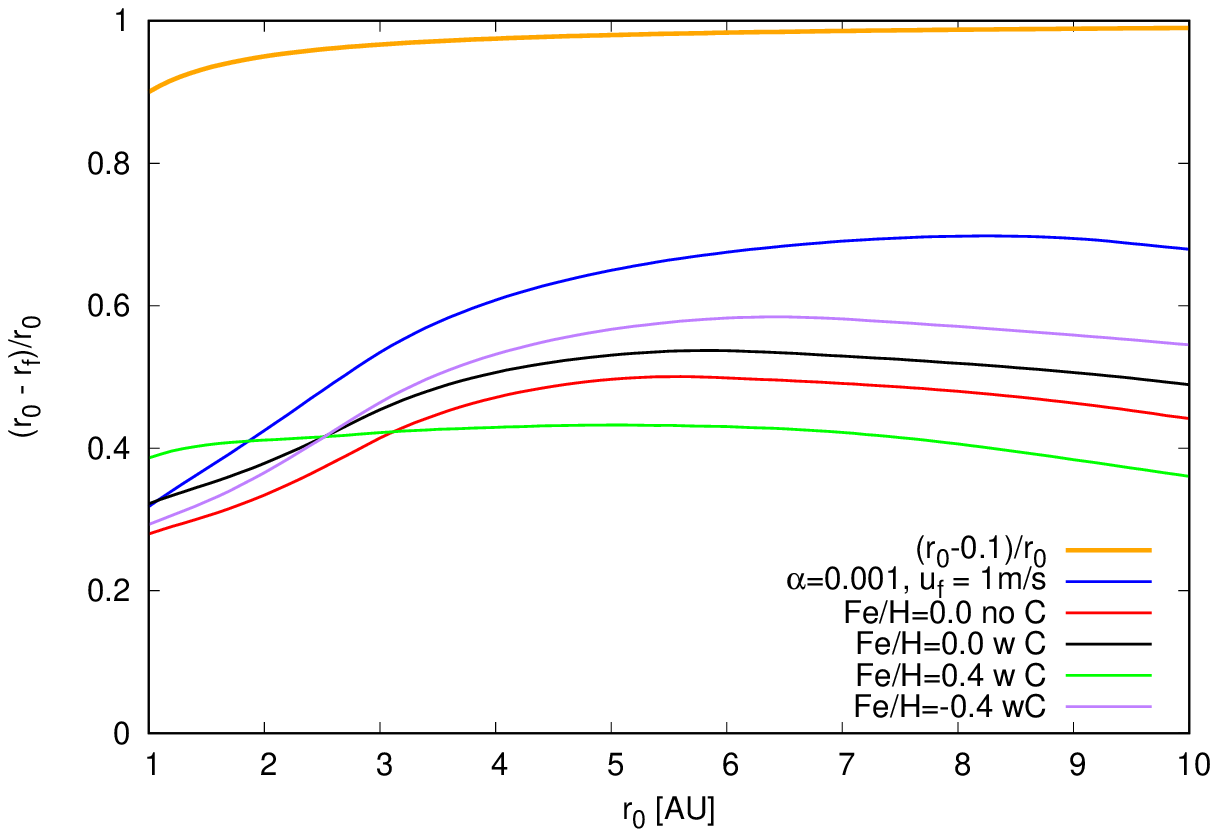}  
 \caption{Change of semi-major axis relative to the initial planetary positions for planetary cores with initially 15 (top) and 25 (bottom) Earth masses in discs with $\alpha=0.001$ and grain size distributions with $u_{\rm f}=1$m/s. The orange line marks the value if the planet reaches the inner edge of the disc at 0.1 AU.
   \label{fig:r0rfMass}
   }
\end{figure}

The gas contraction rates (eq.~\ref{eq:Ikoma2000}) depend crucially on the initial planetary core mass. In fact, the dependency on the planetary core mass is stronger than the dependency on the opacity. We thus show in Fig.~\ref{fig:r0rfMass} the evolution of planetary cores with initially 15 (top) and 25 (bottom) Earth masses. 

As expected, larger planetary cores result in faster gas contraction rates, which results in less inward migration before the runaway phase ($M_{\rm core} = M_{\rm env}$) is reached. For lower core masses, all the planets migrate close to the inner disc edge at 0.1 AU in our simulations. If the core masses are below $\approx$13 Earth masses, all planetary cores migrate to the discs inner edge independently of the opacity. This is caused by the fast inward migration in discs with steep power laws in surface density (eq.~\ref{eq:disc}). This effect is more pronounced in the case of large viscosities, where migration is faster because a partial gap that can slow down migration is prevented to form (eq.~\ref{eq:migII}).

The differences in the final orbital position of the growing planets originating from the different grain compositions are enhanced with increasing planetary core masses. For low core masses these differences become negligible, because the planets migrate all the way to the inner edge before they can reach runaway gas accretion, because the gas contraction rates are too slow due to the low core masses. Only for large core masses (above 15 Earth masses) the compositional differences in opacity seem to play a role for the final planetary positions.

The initial planetary core mass determines if the planet has a chance to reach the runaway gas accretion phase before it migrates to the inner edge, while the chemical composition of the grains that determine the opacity influences the exact final position, if the planet does not migrate to the inner disc edge.

\section{Discussion}
\label{sec:disc}

In this section we discuss the shortcomings and implications of our simulations. 

\subsection{Implications of pebble isolation}

When a planet starts growing above several Earth masses it starts to open up a partial gap in the protoplanetary disc. If the planet continues to grow, it can open a deeper gap that can invert the radial pressure gradient in the protoplanetary disc exterior to the planet \citep{2004A&A...425L...9P, 2006A&A...453.1129P}. If this is the case, inward drifting pebbles will accumulate in this pressure bump and the planet will stop accreting pebbles \citep{2012A&A...546A..18M, 2014A&A...572A..35L, 2018arXiv180102341B, 2018A&A...615A.110A}. This is refered to as the pebble isolation mass.

The pebble isolation mass in itself depends on the properties of the protoplanetary discs, in particular on the aspect ratio, the disc's turbulence viscosity, the underlying radial pressure gradient and also on the Stokes number of the pebbles \citep{2018arXiv180102341B}. This implies that the pebble isolation mass is a function of orbital distance from the star and increases outwards, if H/r increases radially like in a flaring disc model as we use in our simulations. As a consequence, the planetary cores growing by pebble accretion will increase in mass radially \citep{2019A&A...630A..51B}.

In our simulations, however, we used arbitrary pebble isolation masses when envelope contraction starts, not in agreement with the pebble isolation mass in the outer regions in our disc model. Nevertheless what matters are the relative differences arising from the changes in envelope contraction rates originating from the assumptions of fragmentation velocity and $\alpha$ of the grain size distributions and form the different chemical compositions.

As the planet starts to grow, it opens a partial gap in the protoplanetary disc and large pebbles can be blocked exterior to the planetary orbit. However, pebbles that are smaller than the $\alpha$ value of turbulence can drift through the pressure bump \citep{2016A&A...585A..35P, 2018arXiv180102341B, 2020ApJ...896..135C}. In our case, this implies that only a small number of particles are blocked exterior of the planet, even if $\alpha$ is very small (Fig.~\ref{fig:Opacum}). Taking this effect self-consistently into account would reduce the opacity derived in discs with low $\alpha$ by a few percent, thus reducing the envelope opacities and the contraction time, leading to even less inward migration, enhancing the effect that we described. At high viscosities, the Stokes numbers of the particles are so small that all grains would diffuse through the pressure bump (St < $\alpha$). Recently \citet{2020ApJ...896..135C} showed that large pebbles inside the pressure bump can fragment and then diffuse inwards where they increase the opacity inside the planetary envelope, delaying gas accretion. However, the diffusion does not depend on the composition of the grains in the pressure bump. Consequently the enrichment of the planetary envelope with inward diffusing grains is the same independently of their composition, but the underlying opacity differences from the different grain compositions should still matter, even for the then delayed gas accretion.



\subsection{Disc structure}

Our disc model follows a simple power law in gas surface density and temperature that we do not change depending on the grain size distributions and opacities. In real protoplanetary discs, the grain size distribution strongly influences the disc structure \citep{2020arXiv200514097S}. Discs with lower viscosity would also be colder and harbor larger grain sizes that carry less opacity. The simulations by \citet{2016A&A...590A.101B} show that a change in the water-to-silicate ratio in protoplanetary discs influences the disc structure around the water ice line and with this also planet formation. However, we do not take these effects here into account for simplicity. 

The parameters of the disc model strongly influence the migration rates of embedded planets. In particular, the migration rate of embedded planets scales linearly with the disc's surface density. In low density discs, planets would migrate slower. However, the relative differences in the migration speed observed in our simulations would still exist, but might be less or even more dramatic, depending on the underlying disc profile. This highlights the point that planet formation studies need to evolve in a direction that take the influence of the chemical compositions on accretion and migration into account.

\subsection{Implications for planet formation}

Our simulations show that the grain size distribution inside protoplanetary discs does not only influence the disc structure \citep{2020arXiv200514097S} and the solid accretion rates in the pebble accretion scenario\footnote{Pebble accretion depends crucially on the size of the pebbles that are accreted. Larger pebbles result in a larger accretion rate.} \citep{Johansen2017}, but also influences the envelope contraction rates. In the case of larger grains (low viscosity, large fragmentation velocity), the envelope contraction rates increase, leading to a faster growth of planets that then also migrate less before reaching runaway gas accretion.

Our simulations show that the envelope contraction rates are increased in the case of a composition reflecting [Fe/H]=0.4, because of the lower water ice content. This in turn leads to less inward migration during the envelope contraction phase, compared to discs that are water rich (Fig.~\ref{fig:r0rf}). As a consequence, the chemical composition of the protoplanetary disc does not only matter for the final composition of the planet \citep{2020A&A...633A..10B}, but also for the growth rate and migration rate of the planet. Our simulations thus imply that gas giants formed in water poor discs might be located further away from their central star compared to gas giants formed in water rich discs, if their planetary core masses are the same.

In our model we have used a constant dust-to-gas ratio of 0.01, however, giant planets are found around stars that nearly span an order of magnitude in their heavy element content \citep{2004A&A...415.1153S, 2005ApJ...622.1102F, J2010}. The opacity in itself scales linearly with the dust-to-gas ratio, meaning that planets forming in less metal rich environments could contract their envelope faster. On the other hand, the formation of the planetary core is hindered due to the smaller amount of available planetary building blocks, which is reflected in the giant-planet metallicity relation \citep{2004A&A...415.1153S, 2005ApJ...622.1102F, J2010}. Nevertheless, the difference in the chemical composition of the material would still influence planet formation in the same way as described above, even though the absolute envelope contraction rates and migration rates might change depending on the metallicity of the system.




\section{Conclusions}
\label{sec:conclude}

We have studied the influence of grain size distributions derived for environments with different turbulence levels and different assumed grain fragmentation velocities on the contraction rates of planetary envelopes. In addition we have investigated how the chemical composition of the grains change the envelope contraction rates. The compositional changes of the grains and their influence to the opacity seem to have only minimal effects on the migration of non-accreting planets. However, the combined effects of accretion and migration revealed a change in the final semi-major axis of envelope contracting migrating planets with the chemical composition of the grains. Our results are based on the assumption that the grains present in the protoplanetary disc are accreted directly without any further growth and fragmentation in the planetary envelope. However, the grains can evolve inside the envelope \citep{2014ApJ...789L..18O} and also recycling flows inside the planetary Hill sphere could change the envelope contraction rates \citep{2017MNRAS.471.4662C, 2017A&A...606A.146L}, which is not taken into account in our simple model \citep{2000ApJ...537.1013I}.

As expected, grain size distributions leading to larger grains lead to lower opacities, which in turn lead to faster envelope contraction rates and thus to smaller distances the planet migrates until it reaches runaway gas accretion. We summarize our main findings as follows:
\begin{itemize}
 \item[1)] Simplistic envelope opacity assumptions might underestimate the opacity provided by the dust grains accreted into a contracting planetary atmosphere. This could severely shorten the envelope contraction time and thus lead to a significantly smaller distance planets migrate before reaching runaway gas accretion.
 \item[2)] The chemical composition of the dust grains inside the gas that the planet accretes have a strong influence on the envelope contraction rates, where the contraction rates can vary by a factor of a few depending on the chemical composition of the grains. In addition, this then leads to a difference in the distance planets migrate before reaching runaway gas accretion and opening a deep gap, where type-II migration could safe them from migrating all the way to the central star \citep{2017Icar..285..145C}. However, these effects become only important for core masses above 15 Earth masses, because smaller cores contract their envelopes too slow so that all planets migrate to the inner disc edge in our model.
 \item[3)] Our simulations indicate that gas giants forming in water poor environments might be located further away from their central star compared to gas giants forming in water rich environments.
\end{itemize}
We conclude that future simulations aimed to study the formation of planetary systems need to take the effect of grain size distributions and of the chemical composition of the grains for envelope contraction rates and migration rates into account. This becomes particularly important if the formation pathway of individual planetary systems is studied.


\begin{acknowledgements}

B.B. and S.S. thank the European Research Council (ERC Starting Grant 757448-PAMDORA) for their financial support. S.S is a Fellow of the International Max Planck Research School for Astronomy and Cosmic Physics at the University of Heidelberg (IMPRS-HD). We thank an anonymous referee for her/his comments that helped to improve our manuscript.

\end{acknowledgements}

\bibliographystyle{aa}
\bibliography{Stellar}
\end{document}